# Dark states and vortex solutions with finite energy of Maxwell-Dirac nonlinear equations

L. M. Kovachev

*Institute of Electronics, Bulgarian Academy of Sciences, 72 Tzarigradcko shossee, 1784 Sofia, Bulgaria*

**Abstract:** Starting from Maxwell equations for media with no-stationary linear and nonlinear polarization, we obtain a set of nonlinear Maxwell amplitude equations in approximation of first order of the dispersion. After a special kind of complex presentation, the set of amplitude equations was written as a set of nonlinear Dirac equations. For broad-band pulses solitary solutions with half-integer spin and finite energy are found. The solutions correspond to electromagnetic wave with circular Poynting vector and zero divergence. These invisible for detectors waves are called dark states and the localized energy we determine as electromagnetic mass.

1. Intoduction

The problems with the linear equations, describing propagation and interaction of elementary particles was understand in the 1950's-1960's and the investigations was turn on two similar ways. The first one is by direct nonlinear generalization of the Dirac equation [1-5] and other one is by simply adding new (nonlinear) terms to the Lagrangians [6-7]. The second approach was more productive and on this base the cotemporary quantum field theory was born. Naturally, these nonlinear Lagrangians are invariant to some rotational groups, which gives the possibility for classification of the elementary particles [7]. As it is well known, the Dirac equations are derived from the Klein-Gordon wave equation, where the nonlinear generalization request equal nonlinear terms in the four component of the Dirac field. The equality of the all four spinors, was the main problem in the direct nonlinear approach of Dirac equations. This requirement was formally avoided in the second approach from Feynman and Gell-Mann [6] by Lagrangian separation of the Dirac field on left- and right-handed spinors. In this way in the electroweak theory was introduced asymmetry between left- and right-hand weak interaction.

Thus, the natural question arises: Is other possibilities for generation of asymmetries in the nonlinear part of Dirac equations? The answer of this question is in the relationship between equations of the classical electrodynamics and Dirac equations. Particularly, in [8, 9] was shown that the Dirac equations for the neutrino and for the free electron coincide with the Maxwell equations for a suitable current distribution. The results were obtained from mathematical principles, using Clifford algebra formalism and the general relativity theories, but the electrical and magnetic currents are far away from the currents and polarization of a real media. From other hand, the complex structure of the Dirac equations is obtained after presentation of a new vector field as result of mixing of the electrical and magnetic components of the kind $\vec{F} = \vec{E} + i\vec{B}$. In this way the well-known effect as linear and nonlinear polarization and magnetization of the media are mixed and there are no clear physical interpretation in this situation. This is the reason in our investigation to look for new kind of complex presentation, where the linear and nonlinear polarization effects are separated form magnetization of the media [10, 11]. This new kind of complex presentation allow us to obtain two kind of Pauli spinors, where the first one is associated with the first Maxwell equation and the polarization of the electrical field, while the second one is associated with second Maxwell equation and the magnetization of the media. In this way we separate the polarization from magnetic effects in the Dirac form in the same manner as it is in Maxwell equations.

In our investigation we start with the well-known polarization properties of arbitrary isotropic media using in fact that for localized optical waves the response of the media is non-stationary. One appropriate technique for solving the corresponding Maxwell nonlinear integro-differential equations is by using amplitude approximation directly to the Maxwell's system, instead to the integro-differential nonlinear wave equation, as it is performed in the nonlinear optics. We use electrical current, without free charges of a media with linear and nonlinear non-stationary polarization. As a first step, set of nonlinear amplitude Maxwell equations (NME) was obtained. The NMEs admits more reach mathematical structure than the well-known amplitude equations, obtained from the wave equation in approximation of second order linear and zero-order nonlinear dispersion. In particular, they keeps on the difference between the linear $\varepsilon(\omega)$ and nonlinear $\hat{\chi}^{(3)}(\omega)$ dispersion of the media, and the influence of the magnetic permeability $\mu_0$. After a special complex transformation of the amplitude functions of NMEs right way we obtain nonlinear Dirac equations with nonlinear terms in the first Pauli spinor of the Dirac field only [11, 12]. This asymmetry is due of certain physical principles – in arbitrary media, including the nonlinear vacuum, the polarization of the media dominate with 5- 6 order of magnitudes than magnetization. This is the reason to appearing of Kerr type nonlinearities in the first two spinors of the Dirac fields. Using this technique, in our first attempt, we obtain localized analytical solutions of the corresponding Dirac equation with spin $j = \pm 1/2$. One of the problems with these solutions is that the radial parts of them are so called "de Broglie solitons", $A(r) = \sin(\alpha r)/r$, and admit infinite energy. In spite of this, if we perform inverse transformation from Dirac to electromagnetic presentation, the corresponding localized electromagnetic wave admits circular Pointing vector with zero divergence, i.e. some kind of dark state.

In this paper we found new class localized solutions of the nonlinear Dirac equation, with finite energy, with spin $j = \pm 1/2$ and again with circular Pointing vector. The main restriction to obtain these solutions is that they are for broad-band, ultra-short localized waves. As it is well known, the amplitude approximation is correct for phase modulated waves, as well as for spectrally limited waves with one-two oscillating periods into the pulse envelopes.

## 2. Maxwell's equations for media with no stationary linear and nonlinear electric polarization

Consider Maxwell's equations in the cases of a source-free medium with no stationary linear and nonlinear electric polarization and stationary magnetic polarization (the case with dispersion). As it is well known the magnetic polarization in arbitrary media is of 5 to 6 order of magnitudes weak than the electric ones. Thus, the process of propagation of electromagnetic wave can be described by the following integro-differential Maxwell equations [13]:

$$\nabla \times \vec{E} = -\frac{1}{c}\frac{\partial \vec{B}}{\partial t}, \qquad (1)$$

$$\nabla \times \vec{H} = \frac{1}{c}\frac{\partial \vec{D}}{\partial t}, \qquad (2)$$

$$\nabla \cdot \vec{D} = 0, \qquad (3)$$

$$\nabla \cdot \vec{B} = \nabla \cdot \vec{H} = 0, \qquad (4)$$

$$\vec{B} = \mu_0 \vec{H}, \quad \vec{D} = \vec{P}_{lin} + \vec{P}_{nl}, \qquad (5)$$

where $\vec{E}$ and $\vec{H}$ denote the electric and magnetic fields, $\vec{D}$ and $\vec{B}$ refer to the electric and magnetic induction respectively and the constant $\mu_0$ is the magnetic permeability of the dielectric media. In the present paper we will investigate the propagation of optical pulses in isotropic media and the terms of linear $\vec{P}_{lin}$, and nonlinear polarization $\vec{P}_{nl}$ are presented as

$$\vec{P}_{lin} = \int_0^\infty \left[\delta(t-\tau) + 4\pi\tilde{\chi}^{(1)}(\tau)\right]\vec{E}(t-\tau,x,y,z)d\tau = \int_0^\infty \tilde{\varepsilon}^{(1)}(\tau)\vec{E}(t-\tau,x,y,z)d\tau, \qquad (6)$$

$$4\pi\vec{P}_{nl} = 4\pi\int_0^\infty\int_0^\infty\int_0^\infty \tilde{\chi}^{(3)}(\tau_1,\tau_2,\tau_3)\left[\vec{E}(t-\tau_1,r)^* \cdot \vec{E}(t-\tau_2,r)\right]\vec{E}(t-\tau_3,r)d\tau_1 d\tau_2 d\tau_3, \qquad (7)$$

where $\tilde{\chi}^{(1)}(\tau)$, $\tilde{\chi}^{(3)}(\tau_1,\tau_2,\tau_3)$ and $\tilde{\varepsilon}^{(1)}(\tau)$ are the linear and nonlinear electric susceptibility response and the dielectric response respectively. It is more convenient the connected charge density to be calculated with stationary optical response and linear $\varepsilon_0$ and nonlinear $n_2$ constants, as it was perform in [14]. The divergence of the electrical field due to linear and nonlinear polarization is simply

$$\nabla \cdot \vec{E} = \nabla\left[\ln\left(1+\frac{n_2}{\varepsilon_0}|\vec{E}|^2\right)\right]\cdot \vec{E}, \qquad (8)$$

where the expression

$$\rho(x,y,z) = \nabla\left[\ln\left(1+\frac{n_2}{\varepsilon_0}|\vec{E}|^2\right)\right] \cong 0, \quad (9)$$

plays the role of charge density connected with the nonlinear polarization.

## 3. Derivation of the amplitude equations

In this section, we derive the amplitude approximation of Maxwell equations (1-5) (NME) with no-stationary linear and nonlinear electric responses (6, 7) in a standard way [11]. We define the electric and magnetic fields amplitudes by the following relations:

$$\vec{E}(x,y,z,t) = \vec{A}(x,y,z,t)\exp[i(\omega_0 t - g(x,y,z))], \quad (10)$$

$$\vec{H}(x,y,z,t) = \vec{C}(x,y,z,t)\exp[-i(\omega_0 t - q(x,y,z))], \quad (11)$$

where $\vec{A}$ and $\vec{C}$ are the complex amplitudes of the electric and magnetic fields, $\omega_0$, $g$, and $q$ are the optical frequency, and the real spatial phase functions, respectively. By using Fourier representation of the response functions (6), and (7), we obtain the following expressions for the first derivatives in time of the linear and nonlinear electric polarization, and magnetic induction fields

$$\frac{1}{c}\frac{\partial \vec{P}_{lin}}{\partial t} = i\exp(i(\omega_0 t - g(x,y,z)))\left(\int_{-\infty}^{\infty}\frac{\omega\hat{\varepsilon}(\omega)}{c}\vec{A}(x,y,z,\omega-\omega_0)\exp[i(\omega-\omega_0)t]d\omega\right) \quad (12)$$

$$\frac{4\pi}{c}\frac{\partial \vec{P}_{nlin}}{\partial t} = i\exp(i(\omega_0 t - g(x,y,z)))\times$$
$$\iiint_\omega \frac{12\pi\omega\,\hat{\chi}^{(3)}(\omega)}{c}\left|\vec{A}(r,\omega-\omega_0)\right|^2\left(\vec{A}(r,\omega-\omega_0)\exp[i(\omega-\omega_0)t]\right)(d\omega)^3 \quad (11)$$

$$-\frac{1}{c}\frac{\partial \vec{B}}{\partial t} = -i\exp(-i(\omega_0 t - q(x,y,z)))\int_{-\infty}^{\infty}\frac{\omega\mu_0}{c}\vec{C}(x,y,z,\omega-\omega_0)\exp[-i(\omega-\omega_0)t]d\omega, \quad (13)$$

where $\hat{\varepsilon}(\omega)$, $\hat{\chi}^{(3)}(\omega)$ are the linear and nonlinear dispersion of the media. At this point, we restrict the spectrum of the amplitudes of electrical and magnet fields by writing the wave numbers in a Taylor series:

$$k_{lin} = \frac{\omega\hat{\varepsilon}(\omega)}{c} = k_{lin}^0(\omega_0) + \frac{\partial k_{lin}}{\partial\omega}(\omega-\omega_0) + \ldots = k_{lin}^0(\omega_0) + \frac{1}{v_{gr}}(\omega-\omega_0) + \ldots, \quad (14)$$

$$k_{nlin} = \frac{12\pi\omega\hat{\chi}^{(3)}(\omega)}{c} = k_{nlin}^0(\omega_0) + \frac{\partial k_{nlin}}{\partial\omega}(\omega-\omega_0) + \ldots, \quad (15)$$

$$k_{mag} = \frac{\omega\mu_0}{c} = k_{mag}^0(\omega_0) + \frac{\mu_0}{c}(\omega-\omega_0) + \ldots \quad (16)$$

The nonlinear wave number can be expressed as

$$k_{nlin} = \frac{12\pi\omega_0 \hat{\chi}^{(3)}(\omega)}{c} = \frac{\omega_0 \hat{\varepsilon}_0}{c} \frac{12\pi\hat{\chi}^{(3)}(\omega)}{\hat{\varepsilon}_0} = k_{lin} n_2, \tag{17}$$

where $v_{gr}$ have dimension of group velocity, and

$$n_2 = \frac{12\pi\hat{\chi}^{(3)}(\omega)}{\hat{\varepsilon}_0}, \tag{18}$$

is the nonlinear refractive index of the media. It is important to be pointed that the series are strongly convergent for pulses up to single cycle regime (broad-band waves). After a couple of transformations the expressions of the first derivatives are presented in the form

$$\frac{1}{c}\frac{\partial \vec{P}_{lin}}{\partial t} = \exp(i(\omega_0 t - g(x,y,z)))\left( ik_{lin}^0 \vec{A} + \frac{1}{v_{gr}}\frac{\partial \vec{A}}{\partial t} + ....\right), \tag{20}$$

$$\frac{4\pi}{c}\frac{\partial \vec{P}_{nlin}}{\partial t} = \exp(i(\omega_0 t - g(x,y,z)))\left( ik_{lin}^0 n_2 |\vec{A}|^2 \vec{A} + ...\right), \tag{21}$$

$$-\frac{1}{c}\frac{\partial \vec{B}}{\partial t} = -\exp(-i(\omega_0 t - q(x,y,z)))\left( ik_{mag}^0 \vec{C} + \frac{\mu_0}{c}\frac{\partial \vec{C}}{\partial t} + ....\right). \tag{22}$$

As it can be seen in the next paragraph, we investigate real in space vector functions, multiplied with the complex in time variables. For such spatial vector field the divergence of the amplitude functions can be written in the form

$$\nabla \cdot \vec{A} = \left\{\nabla\left[\ln\left(1 + \frac{n_2}{\varepsilon_0}|\vec{A}|^2\right)\right] - \tan(g(r))\nabla g(r)\right\} \cdot \vec{A} \cong \left\{\nabla\left[\ln\left(1 + \frac{n_2}{\varepsilon_0}|\vec{A}|^2\right)\right] - r\nabla g(r)\right\} \cdot \vec{A} \approx 0, (23)$$

up to one small but finite value of $r = r_0$. Finally, from the Maxwell's equations we obtain the following set of nonlinear amplitude equations

$$\nabla \times \vec{A} = -\left(ik_{mag}^0 \vec{C} + \frac{\mu_0}{c}\frac{\partial \vec{C}}{\partial t} + \right)..., \tag{24}$$

$$\nabla \times \vec{C} = ik_{lin}^0 \vec{A} + \frac{1}{v_{gr}}\frac{\partial \vec{A}}{\partial t} + ik_{lin}^0 n_2 |\vec{A}|^2 \vec{A} + .... \tag{25}$$

$$\nabla \cdot \vec{A} = 0, \tag{26}$$

$$\nabla \cdot \vec{C} = 0, \tag{27}$$

if the gradient of the real spatial phase functions $g$ and $q$ satisfy the relations

$$\nabla g \times \vec{A} = 0, \text{ and } \nabla q \times \vec{C} = 0, \tag{28}$$

**4. Complex representation and Dirac form of NMEs**

To solve the NME Eqs. (24) - (27) we look for stationary solution and as a step we represent the complex amplitude functions as

$$\vec{A}(x, y, z, t) = \vec{F}(x, y, z)\exp(-i\Delta\omega t), \tag{29}$$

$$\vec{C}(x, y, z, t) = \vec{G}(x, y, z)\exp(i\Delta\omega t). \tag{30}$$

Substituting these forms into NMEs (24)-(27) we obtain

$$\nabla \times \vec{F} = -i\kappa_{mag}\vec{G} \tag{31}$$

$$\nabla \times \vec{G} = i\kappa_{opt}\vec{F} + i\gamma_2|\vec{F}|^2\vec{F} + .... \tag{32}$$

$$\nabla \cdot \vec{F} = 0, \tag{33}$$

$$\nabla \cdot \vec{G} = 0, \tag{34}$$

where the generalized wave numbers takes the form

$$\kappa_{opt} = k_{lin} - \Delta k = \frac{\omega_0 \varepsilon(\omega_0)}{c} - \frac{\Delta\omega}{v_{gr}} \tag{35}$$

$$\kappa_{mag} = k_{mag}^0 + \Delta k = \frac{\mu(\omega_0)}{c}(\omega_0 + \Delta\omega) \tag{36}$$

and $\gamma_1 = n_2 k_{lin}^0$. As it well known, for phase modulated broadband waves the spectral bandwidths $\Delta k$ can reach the values $\Delta k \cong k_{lin}$ and then the generalized optical wave number vanish $\kappa_{opt} \cong 0$. Let's now perform the complex presentation of the NMEs (31)-(34) by substitutions

$$\Psi_1 = iF_z, \tag{37}$$

$$\Psi_2 = iF_x - F_y, \tag{38}$$

$$\Psi_3 = -G_z, \tag{39}$$

$$\Psi_4 = -iG_y - G_x. \tag{40}$$

Thus, we obtain the following nonlinear Dirac system of equations (NDE)

$$\left(\frac{\partial}{\partial x} - i\frac{\partial}{\partial y}\right)\psi_4 + \frac{\partial}{\partial z}\psi_3 = -i\left(\kappa_{opt} + \gamma_1\sum_{i=1}^{2}|\Psi_i|^2\right)\Psi_1 \tag{41}$$

$$\left(\frac{\partial}{\partial x} + i\frac{\partial}{\partial y}\right)\psi_3 - \frac{\partial}{\partial z}\psi_4 = -i\left(\kappa_{opt} + \gamma_1\sum_{i=1}^{2}|\Psi_i|^2\right)\Psi_2 \tag{42}$$

$$\left(\frac{\partial}{\partial x}-i\frac{\partial}{\partial y}\right)\psi_2+\frac{\partial}{\partial z}\psi_1=-i\kappa_{mag}\Psi_3 \tag{43}$$

$$\left(\frac{\partial}{\partial x}+i\frac{\partial}{\partial y}\right)\psi_1-\frac{\partial}{\partial z}\psi_2=-i\kappa_{mag}\Psi_4 \tag{44}$$

Note, that the optical NDEs are significantly different from the NDEs in the field theory. The nonlinear part appears ONLY in the first two coupled equations of the system. Other important result is that the first two equations in the Dirac system correspond to first Maxwell equation, while the third and four equations in the system are connected with the second Maxwell equation. The divergent terms are incorporate in the system also.

**5. Broadband finite energy stationary vortex solutions of NDEs with spin $j=\pm 1/2$**

In two previous papers [11, 12] we investigate the NDEs for narrow-band optical waves where the relation for the optical wave number is $\kappa_{opt}=k_{lin}-\Delta k>0$. For this types optical waves we found vortex solutions with spin $j=\pm 1/2$ and radial part, proportional to "de-Broglie" soliton solutions of the types $\eta(r)=\sin(r)/r$. It is well known that such type localizations admit infinite integral of energy. That why we start to look for solutions in the case of broad-band waves, when

$$\kappa_{opt}=k_{lin}-\Delta k=\frac{\omega_0\varepsilon(\omega_0)}{c}-\frac{\Delta\omega}{v_{gr}}\cong 0; \quad \kappa_{mag}=\frac{2\mu(\omega_0)\omega_0}{c} \tag{45}$$

In this case the NDEs is transformed to

$$\left(\frac{\partial}{\partial x}-i\frac{\partial}{\partial y}\right)\psi_4+\frac{\partial}{\partial z}\psi_3=-i\left(\gamma_1\sum_{i=1}^{2}|\Psi_i|^2\right)\Psi_1 \tag{46}$$

$$\left(\frac{\partial}{\partial x}+i\frac{\partial}{\partial y}\right)\psi_3-\frac{\partial}{\partial z}\psi_4=-i\left(\gamma_1\sum_{i=1}^{2}|\Psi_i|^2\right)\Psi_2 \tag{47}$$

$$\left(\frac{\partial}{\partial x}-i\frac{\partial}{\partial y}\right)\psi_2+\frac{\partial}{\partial z}\psi_1=-i\kappa_{mag}\Psi_3 \tag{48}$$

$$\left(\frac{\partial}{\partial x}+i\frac{\partial}{\partial y}\right)\psi_1-\frac{\partial}{\partial z}\psi_2=-i\kappa_{mag}\Psi_4 \tag{49}$$

We use the Pauli matrices to rewrite the system Eqs. (46) - (49) in form

$$(\vec{\sigma}\cdot\vec{P})\varphi=-i\left(\gamma_1\sum_{i=1}^{2}|\eta_i|^2\right)\eta, \tag{50}$$

$$(\vec{\sigma} \cdot \vec{P})\eta = -i\kappa_{mag}\varphi, \tag{51}$$

where $\vec{\sigma}$ are the Pauli matrices, $\vec{P} = (\partial/\partial x, \partial/\partial y, \partial/\partial z)$ is the differential operator and

$$\eta = \begin{pmatrix} \Psi_1 \\ \Psi_2 \end{pmatrix}; \quad \varphi = \begin{pmatrix} \Psi_3 \\ \Psi_4 \end{pmatrix},$$

are the two components spinors. After substituting Eq. (51) into Eq. (50) we obtain for the case, when there is no external electric or magnetic fields

$$\Delta\eta + \kappa_{mag}\gamma_1 \sum_{i=1}^{2} |\eta_i|^2 \eta = 0, \tag{52}$$

where $\Delta$ is the Laplacian operator.

We look for spinors in the following two forms

$$\eta = \begin{pmatrix} \rho(r)\cos(\theta/2) \\ \rho(r)\sin(\theta/2)\exp(i\varphi/2) \end{pmatrix}, \text{ or } \eta = \begin{pmatrix} \rho(r)\sin(\theta/2)\exp(-i\varphi/2) \\ -\rho(r)\cos(\theta/2) \end{pmatrix}. \tag{53}$$

This corresponds to spinors with opposite spin $j = \pm 1/2$. After substituting Eq. (53) into Eq.(52) the following equation for the radial part is obtained

$$\frac{2}{r}\frac{\partial\rho}{\partial r} + \frac{\partial^2\rho}{\partial r^2} - \frac{2}{r^2}\rho + \kappa_{mag}\gamma_1 |\rho|^2 \rho = 0. \tag{54}$$

We rewrite this equation in dimensionless form by using the expressions $r' = r_0 r$ and $\rho' = A_0\rho$. Thus, dimensionless equation becomes (the primes are omitted for clarity)

$$\frac{2}{r}\frac{\partial\rho}{\partial r} + \frac{\partial^2\rho}{\partial r^2} - \frac{2}{r^2}\rho + \tilde{\gamma}_1 |\rho|^2 \rho = 0, \tag{55}$$

where $\tilde{\gamma}_1 = n_2 A_0^2 \kappa_{mag} k_{lin} r_0^2$ is the dimensionless nonlinear constant.

Eq. (47) admits the algebraic solitary solution of the kind

$$\rho(r) = \sec h(\ln r^\alpha)/r, \tag{56}$$

if the following linear and nonlinear dispersion relations are satisfied

$$\alpha^2 = 2 \text{ or } \alpha = \sqrt{2} \text{ and } \tilde{\gamma}_1 = n_2 A_0^2 \kappa_{mag} k_{lin} r_0^2 = 2\alpha^2 = 4. \tag{57}$$

After solving the nonlinear part of the NDEs for linear spinor we have simply

$$\tilde{\phi}(r) = \left(\frac{\partial\rho}{\partial r} + \frac{2}{r}\rho\right)/\kappa_{mag}, \tag{58}$$

where $\tilde{\varphi}$ is the normed radial component of the second spinor.

Substituting the solution (56) into Eq. (58), we obtain solitary solution for the magnetic part of NDEs.

$$\tilde{\varphi}(r) = -\frac{1}{\kappa_{mag}}\left[\alpha \sec h(\ln(r^\alpha))th(\ln(r^\alpha)) + \sec h(\ln(r^\alpha))\right]r^{-2}\ ;\ \alpha = \sqrt{2} \tag{59}$$

Finally the soliton solutions with spin $j = \pm 1/2$ takes the form:

For spin $j = 1/2$

$$\Psi_1 = \rho(r)\cos(\theta/2);\quad \Psi_2 = \rho(r)\sin(\theta/2)\exp(i\varphi/2);\quad \Psi_3 = -i\tilde{\phi}(r);\quad \Psi_4 = 0 \tag{60}$$

For spin $j = -1/2$

$$\Psi_1 = \rho(r)\sin(\theta/2)\exp(i\varphi/2);\quad \Psi_2 = -\rho(r)\cos(\theta/2);\quad \Psi_3 = 0;\quad \Psi_4 = -i\tilde{\phi}(r) \tag{61}$$

The radial part of the first spinor Eq. (56) can be written in the simplest form

$$\rho(r) = \frac{2r^{(\sqrt{2}-1)}}{\left(r^{2\sqrt{2}}+1\right)} \cong \frac{2r^{0.4142}}{\left(r^{2.8284}+1\right)} \tag{62}$$

In the nominator we have dependence with positive degree of the kind $r^{0.4142}$ which is multiplied with a Lorentzian function. Thus, the solitary solution Eq.(62), as well as the intensity of the electrical field, connected with the first two component of the spinor field $\left(I(r) = \sum_{i=1,2}|\Psi_i|^2 = \rho^2(r)\right)$, in the origin tends to zero, while in infinity vanish again due to Lorentzian shape (Fig=1). The same is for magnetic solution Eq. (59) as a derivative of smooth localized function.

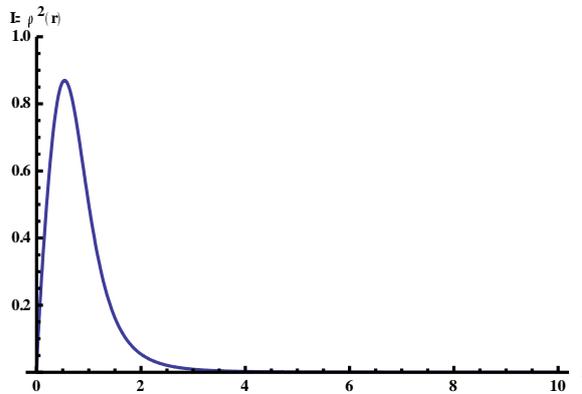

Fig.1. Distribution of the intensity profile of our spinor field. The intensity of the electrical field, connected with the first two component of the spinor field $I(r) = \sum_{i=1,2} |\Psi_i|^2 = \rho^2(r)$, in the origin tends to zero, while in infinity vanish again due to Lorentzian shape.

## 6. Spatial phase functions, Poynting vector and flow of energy

To understand the real dynamics of these solutions, we must turn back from the complex NDEs form to real electromagnetic field in NMEs. Using again relations between the spinors in NDEs and amplitude functions of NMEs we have

$$F_z = \frac{\Psi_1 - \Psi_1^*}{2i}; \quad F_y = \frac{\Psi_2 + \Psi_2^*}{2}; \quad F_x = \frac{\Psi_2 - \Psi_2^*}{2i} \tag{63}$$

$$G_z = -\frac{\Psi_3 - \Psi_3^*}{2i}; \quad G_y = -\frac{\Psi_4 - \Psi_4^*}{2i}; \quad F_x = -\frac{\Psi_4 + \Psi_4^*}{2} \tag{64}$$

Substituting the solutions Eqs. (60) spin $j = 1/2$ in Eqs. (63)-(64) we obtain for the real amplitudes of the electromagnetic field

$$F_z = \rho(r)\frac{z}{r}; \quad F_y = \rho(r)\frac{y}{r}; \quad F_x = \rho(r)\frac{x}{r}, \tag{65}$$

$$G_x = 0; \quad G_y = 0; \quad G_z = \tilde{\varphi}(r). \tag{66}$$

Let's now rewrite again the conditions of the spatial phase functions Eq. (28)

$$\nabla g \times \vec{A} = 0, \text{ and } \nabla q \times \vec{C} = 0,$$

where

$$\vec{A}(x,y,z,t) = \vec{F}(x,y,z)\exp(-i\Delta\omega t), \quad \vec{C}(x,y,z,t) = \vec{G}(x,y,z)\exp(i\Delta\omega t). \tag{67}$$

The phase modulated spatial phase functions of the electrical and magnetic fields, which satisfy the above relations are of the kind

$$g = k_0(x^2 + y^2 + z^2), \tag{68}$$

$$q = k_0 z^2, \quad g - q = k_0(x^2 + y^2). \tag{69}$$

The Poynting vector can be expressed by the solutions for the electromagnetic field in approximation to minimal divergence of the electrical field (minimal polarization charge)

$$\vec{S}(x,y,z,t) = \vec{E} \times \vec{H} = \vec{F}(x,y,z) \times \vec{G}(x,y,z)\exp[i(2\omega_0 t - (g-q))]. \tag{70}$$

Substituting the solutions of the amplitude functions Eqs. (60) in Eq. (70), and using in fact the difference between the spatial phase functions of electrical and magnetic fields we have

$$\vec{S}(x,y,z,t) = \vec{E} \times \vec{H} = \left[ -\rho(r)\tilde{\phi}(r)\frac{y}{r}, \rho(r)\tilde{\phi}(r)\frac{x}{r}, 0 \right] \exp\left[ i\left(2\omega_0 t - k_0\left(x^2 + y^2\right)\right) \right] \quad (71)$$

Finally, using divergence of the Poynting vector $\nabla \cdot \vec{S}$ we obtain the remarkable result, that for solutions with spin $j = 1/2$ the divergence vanish

$$\nabla \cdot \vec{S} = 0. \quad (72)$$

The relation for zero divergence of $\vec{S}$ - Eq. (72), determines that the Poynting vector of our solitary solutions with spin $j = 1/2$ (60) is circular ones and no flow of energy thought surface around the solutions. More precisely, expression for the Poynting vector Eq. (71) determines that the flow of energy oscillate radially in $(x, y)$ plane, while the magnetic field oscillate in direction orthogonal to this plane.

This is the reason our solutions for the localized electromagnetic field to be called dark states. The impossibility for measuring with electromagnetic detectors the flow of energy through a surface determines this localized energy as electromagnetic mass of these localized solitary solutions. The radial shape of the amplitude functions, vanishing in the origin and infinity and show that they have toroid structure.

**Conclusion**

Up to now, the standard way for complex presentation of the Maxwell equation into the Dirac form is by mixing of the electrical and magnetic components of the kind $\vec{F} = \vec{E} + i\vec{B}$. In this complex structure the well-known effect as linear and nonlinear polarization and magnetization of the media are mixed and there are no clear physical interpretation of the results.

In this paper we use a new kind of complex presentation, where the electrical and magnetic components are separated in the Dirac presentation [10, 11]. This allow us to obtain two kind of Pauli spinors, the first one associated with the first Maxwell equation and the linear and nonlinear polarization of the electrical field, and the second one associated with second Maxwell equation and the magnetization of the media. This presentation preserve the separation of polarization from the magnetic effects in the Dirac form of Maxwell equations. Other physical requirement that we use following from the fact that for localized optical waves the response of the media is non-stationary. That why we investigate the corresponding Maxwell nonlinear integro-differential equations is by using amplitude approximation directly to the Maxwell's system, instead to the integro-differential nonlinear wave equation, as it is performed in the nonlinear optics. In first approximation of dispersion set of nonlinear amplitude NME was obtained. After a special complex transformation of the amplitude functions of NMEs, we obtain nonlinear Dirac equations with nonlinear terms in first Pauli spinor of the Dirac field only. This allow us, using the well-known from the quantum mechanics technique, to find localized solutions of the NDEs with finite energy and spin $j = \pm 1/2$. The radial part of the solution is a Lorentzian type functions, vanishing in the origin and infinity and look like as toroid. To understand the real electromagnetic structure of this localized solution, we perform inverse transformation from Dirac to Maxwell form. We obtain that while the corresponding electrical field oscillate in radial direction, the magnetic field admits one component and oscillate in one (*z* for example) direction only.

Other main result is that that for solutions with spin $j = 1/2$ the divergence of the Poynting vector for the electromagnetic field vanish ($\nabla \cdot \vec{S} = 0$) and this is the reason our solutions to be called dark states. The impossibility for measuring with electromagnetic detectors the flow of energy through a surface determines this localized energy as electromagnetic mass of these localized solitary solutions.

The fact that the electrical field of these dark state oscillate and is phase modulated in a radial direction request spherical resonators to be used in a future experiments.

**Acknowledgements:** This work was supported by the National Science Fund (Bulgaria) [grant DN 11/18].

**References:**

1. W. Heisenberg, "Zur Quantentheorie nichtrenormierbarer Wellengleichungen", Zs. Naturforsch., **9a**, 292-303 (1954).

2. W. Heisenberg , "Zur Quantentheorie nichtlinearer Wellengleichungen III", Zs. Naturforsch., **10a**, 425-446 (1955).

3. R. Finkelstein, C. Fronsdal and P. Kaus, Phys. Rev. **103**, 1571-1579 (1956).

4. R. Ascoli und W. Heizenberg, "Zur Quantentheorie nicht linearer Wellengleichungen. IV. Elektrodynamik.", Zs. Naturforsch., **12a,** 177-187 (1957).

5. W. Heisenberg, "Quantum theory of fields and elementary particles." Reviews of modern physics, **29**, 269—278 (1957).

6. R. P. Feynman and M. Gell-Mann, "Theory of the Fermi interaction" Phys. Rev., **109**, 193-198(1958).

7. W. Heisenbrg, H. Dûrr, H. Mitter, S. Schlieder and K. Yamazaki, " "Zur Theorie der Elementarteilchen." Zs. Naturforsch., **14a**, 441—485. (1959).

8. Alfonso Campolattaro, "Classical electrodynamics and relativistic quantum Mechanics", Advances in Applied Cliford Algebra, **7(S)**, 167-173 (1997).

9. Vaz Jame Jr. and Waldyr A. Rodrigues Jr., "Equivalence of Dirac and Maxwell equations and quantum mechanics", Int. J. of Theor. Phys., **32(6)**, 945-959 (1993).

10. A. Dacev, *Quantum mechanics*, Nauka i Izkustvo, Sofia, 1973.

11. L. M. Kovachev, "Optical leptons", International Journal of Mathematics and Mathematical Sciences, **24**, pp.1403-1422 (2004).

12. L. M. Kovachev, "Vortex solutions of the Nonlinear Optical Maxwell-Dirac Equations", Physica D-Nonlinear Phenomena, **190**, pp. 78-92 (2004).

13. Y. R. Shen, *The principles of nonlinear optics*, Jon Wiley &Sons, 1984.

14. D. R. Andersen, L. M. Kovachev, "Interaction of coupled-vector optical vortices", JOSA B, **19**, 376-384 (2002).